\documentclass[aps,prb,showpacs,twocolumn]{revtex4}
\usepackage{graphicx}
\usepackage{epsfig}
\usepackage{bm}

\begin{document}

\title{Spin relaxation and combined resonance in two-dimensional \\
electron systems with spin-orbit disorder}
\author {V. K. Dugaev$^{1,2}$, E. Ya. Sherman$^{3,4}$, V. I. Ivanov$^{5}$, and J.
Barna\'s$^{6,7}$}

\affiliation{$^1$Department of Physics, Rzesz\'ow University of Technology,
Powsta\'nc\'ow Warszawy 6, 35-959 Rzesz\'ow, Poland \\
$^2$Department of Physics and CFIF, Instituto Superior T\'ecnico, TU Lisbon,
Av. Rovisco Pais 1049-001 Lisbon, Portugal \\
$^3$ Basque Foundation for Science IKERBASQUE, Alameda Urquijo 36-5,
48011, Bilbao, Bizkaia, Spain \\
$^4$Department of Physical Chemistry, Universidad del Pa\'is
Vasco, Bilbao, 48080 Bizkaia, Spain \\
$^5$Insitute for Problems of Materials Science, Ukrainian Academy of
Sciences, Vilde 5, 58001 Chernovtsy, Ukraine \\
$^6$Institute of Molecular Physics, Polish Academy of Sciences,
Smoluchowskiego 17, 60-179 Pozna\'n, Poland \\
$^7$ Department of Physics, Adam
Mickiewicz University, Umultowska 85, 61-614 Pozna\'n, Poland
}

\date{\today}

\begin{abstract}
Disorder in spin-orbit (SO) coupling is an important feature of
real low-dimensional electron structures. We study spin relaxation
due to such a disorder as well as resulting abilities of spin
manipulation. The spin relaxation reveals quantum effects when the
spatial scale of the randomness is smaller than the electron
wavelength. Due to the disorder in SO coupling, a time-dependent 
external electric field generates a spatially random 
spin-dependent perturbation. The resulting electric dipole 
spin resonance in a two-dimensional 
electron gas leads to spin injection
in a frequency range of the order of the Fermi energy.
These effects can be important
for possible applications in spintronics.
\end{abstract}


\maketitle

Electron dynamics in low-dimensional semiconductor structures reveals
features of a spin dependent transport that are interesting for 
fundamental and applied research \cite{Fabian04}. One of the main
ingredients necessary to generate spin dependent
transport in nonmagnetic semiconductor systems is the SO
interaction. Such an interaction offers a possibility of an
efficient and fast spin manipulation with electric fields, which in
turn allows to prepare a required spin
state.\cite{Rashba62,Rashba03,Duckheim08,Khomitsky09,Nowack07,Pioro08,Shen07}
At the same time, spin relaxation and decoherence due to the
SO coupling prevent long-distance spin propagation. Two
models are widely used to describe the SO coupling in
low-dimensional structures: the Rashba and the Dresselhaus ones.
In both models, the SO field and the corresponding spin
precession rate are approximately linear in the electron momentum.
Random evolution in the momentum due to collisions with
impurities, phonons, and other electrons results in randomness in
the spin precession, and thus leads to spin relaxation. However,
in reality both interactions have an intrinsic randomness due to
system imperfections, including the fluctuations in the dopant ion
density \cite{Sherman03,Glazov05} or random bonds at the quantum well (QW)
interface \cite{Golub04}. Even if the mean values of the
Rashba and Dresselhaus fields vanish, their fluctuations
remain and can cause interesting consequences, including memory
effects \cite{Glazov05}, spin  Hall effect  in the finite-size
systems \cite{Moca08}, and spin-dependent
localization.\cite{Tserkovnyak09}

There are at least four different two-dimensional (2D) systems,
where the SO disorder plays an important or crucial role. First,
the effect of random SO coupling can be responsible for the spin
relaxation in Si/Ge QWs. \cite{Jantsch02,Golub04}
Second, the spin-dependent disorder influences \cite{Liu06} the
spin helix pattern recently observed in the GaAs (001) QW 
with the balanced Rashba and Dresselhaus
terms.\cite{Koralek09} Third, 
the randomness causes relaxation of
the spin component along the growth axis observed in Ref.[\onlinecite{Muller08}]
in GaAs (011) QW, investigated now 
for spintronics applications.\cite{Muller08,Belkov08} Fourth,
the most recent example of the system with random SO coupling is
graphene, where the randomness and spin relaxation appear due to
the rippling of the layers \cite{Brataas07} and due to the
disorder and electron-phonon coupling in the substrate.\cite{Ertler09}

In this paper we study the effects of randomness on the spin
relaxation and spin injection. We show that
spin relaxation reveals interesting quantum effects arising
from the non-commutativity of the momentum and
coordinate-dependent randomness. The calculated spin injection 
can be observed in a wide range of
frequencies, extended up to the electron Fermi energy.

{\it Model.} We consider a two-dimensional electron gas with
fluctuating Rashba SO interaction. In the absence of external
fields, the Hamiltonian has the form (we
use units with $\hbar =1$) $H=H_{0}+H_{\mathrm{so}}$, where
\begin{eqnarray}
&&H_{0}=-\frac{\nabla ^{2}}{2m}+U(\mathbf{r}), \\
&&H_{\mathrm{so}}=-\frac{i}{2}\sigma _{x}\left\{ \nabla _{y},\, \lambda (%
\mathbf{r})\right\} +\frac{i}{2}\sigma _{y}\left\{ \nabla _{x},\, \lambda (%
\mathbf{r})\right\}.
\label{1} 
\end{eqnarray}
Here $m$ is the electron effective mass, $U(\mathbf{r})$ is the
random potential leading to the momentum relaxation time $\tau_p$, and
$\left\{ ,\right\} $ stands for the anticommutator.  The random Rashba field
$\lambda(\mathbf{r})$ has zero expectation value $\left\langle\lambda (\mathbf{r})\right\rangle =0$ 
and correlation function 
\begin{equation}
C_{ \lambda \lambda }\left(\mathbf{r-r}^{\prime }\right)
\equiv
\langle  \lambda (\mathbf{r})\, \lambda (\mathbf{r}^{\prime })\rangle
=\left\langle \lambda^{2}\right\rangle F\left(\mathbf{r-r}^{\prime}\right).
\label{4}
\end{equation}
The brackets $\left\langle ...\right\rangle $
stands for the average over the disorder, and the range function
$F\left( \mathbf{r-r}^{\prime }\right) $ depends on the disorder
type.

{\it Spin relaxation.} We begin with calculation of spin
relaxation rate due to random $H_{\mathrm{so}}$ interaction.
The eigenfunctions of Hamiltonian $H_{0}$ (normalized to the unit
area) are
$\psi_{\overline{\mathbf{k}}}=e^{i\mathbf{k}\cdot \mathbf{r}}\;\chi _{\sigma }$,
where we included the spin index $\sigma$ in the definition of
momentum $\overline{\mathbf{k}}=\left(
\mathbf{k,}\sigma \right)$, and $\chi _{\sigma }$ is the spin
function. Matrix elements of SO interaction 
$V_{\overline{\mathbf{k}}\overline{\mathbf{k}}\mathbf{^{\prime }}}
\equiv
\langle\overline{\mathbf{k}}|H_{\rm so}|\overline{\mathbf{k}}^{\prime}\rangle$
are
\begin{equation}
V_{\overline{\mathbf{k}}\overline{\mathbf{k}}\mathbf{^{\prime }}}
=\frac{ \lambda_{\mathbf{k}-\mathbf{k}^{\prime }}}{2}
\left<\sigma\right.\left|
\sigma _{x}\left( k_{y}+k_{y}^{\prime }\right)
-\sigma _{y}\left(k_{x}+k_{x}^{\prime }\right)
\right|
\left.\sigma^{\prime}\right>,
\label{5}
\end{equation}
where $ \lambda_{\mathbf{k-k}^{\prime}}$ is the Fourier component
of the random Rashba field. 

\begin{figure}
\includegraphics[scale=0.55]{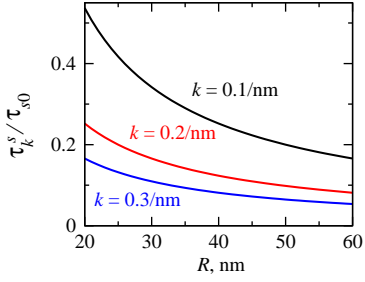}
\caption{(Color online) Spin relaxation time as a
function of the correlation radius $R$ for different electron wave
vectors $k$ marked near the lines.}
\end{figure}

To calculate the spin relaxation time,
we use the kinetic equation for spin density matrix
$\rho_{\overline{\mathbf{k}}}$ (see for instance
Refs.~[\onlinecite{Tarasenko06,Culcer07,Glazov07}])
\begin{equation}
\frac{\partial \rho _{\overline{\mathbf{k}}}}{\partial t}+
i[H_{\mathrm{so}},\,\rho _{\overline{\mathbf{k}}}]=
\mathrm{St}\,\rho_{\overline{\mathbf{k}}}. 
\label{6}
\end{equation}
Here, due to the absence of the regular contribution 
in the $H_{\mathrm{so}}$, the commutator term 
in Eq.(\ref{6}) vanishes. Therefore, the entire effect of SO randomness
is included in the collision integral  \cite{Tarasenko06}
\begin{eqnarray}
\mathrm{St}\,\rho _{\overline{\mathbf{k}}}
=\pi \sum_{\overline{\mathbf{k}}\mathbf{^{\prime }}}
\left(
2V_{\overline{\mathbf{k}}\overline{\mathbf{k}}\mathbf{^{\prime}}}
\rho_{\overline{\mathbf{k}}\mathbf{^{\prime }}}
V_{\overline{\mathbf{k}}\mathbf{^{\prime }}\overline{\mathbf{k}}}
-
V_{\overline{\mathbf{k}}\overline{\mathbf{k}}\mathbf{^{\prime }}}
V_{\overline{\mathbf{k}}\overline{\mathbf{k}}\mathbf{^{\prime }}}
\rho _{\overline{\mathbf{k}}}
-
\rho_{\overline{\mathbf{k}}}
V_{\overline{\mathbf{k}}\overline{\mathbf{k}}\mathbf{^{\prime }}}
V_{\overline{\mathbf{k}}\mathbf{^{\prime }}\overline{\mathbf{k}}}\right)
\nonumber  \\
\times \delta
\left( \varepsilon_{\mathbf{k}}-\varepsilon_{\mathbf{k^{\prime }}}
\right),
\label{7}
\end{eqnarray}
with kinetic energy $\varepsilon_{\mathbf{k}}=k^{2}/2m.$ 
We take $\rho_{\overline{\mathbf{k}}}$ in the form corresponding to the
only nonzero $z$-spin component 
$\rho_{\overline{\mathbf{k}}}=
\rho _{0\overline{\mathbf{k}}}
+S_{\mathbf{k}}\,\sigma _{z}$,
with  $\rho _{0\overline{\mathbf{k}}}$ being the equilibrium
density matrix. The resulting macroscopic spin density: 
\begin{equation}
\left< s_z \right> =\frac{1}{2} \int S_{\mathbf{k}} \frac{d^2k}{(2\pi)^2}.
\label{spin_density}
\end{equation}
 Using Eqs.~(\ref{5}) and (\ref{7}) we find
\begin{eqnarray}
\mathrm{St}\,\rho _{\overline{\mathbf{k}}}=
-\frac{\pi m\sigma _z}{2k}
\sum_{\mathbf{q}}C_{ \lambda  \lambda }(\mathbf{q})
\frac{4k^{2}-q^{2}}{q}
\left( S_{\mathbf{k^{\prime }}}+S_{\mathbf{k}}\right)
\nonumber \\ \times
\delta \left(\frac{q}{2k}-\cos\varphi\right) ,
\label{10}
\end{eqnarray}
where $C_{ \lambda  \lambda }(\mathbf{q})$ is the Fourier transform of the
correlator $C_{ \lambda  \lambda }\left( \mathbf{r}\right) ,$ $\varphi $ is
the angle between $\mathbf{k}$ and $\mathbf{q}$, and $\mathbf{q}=\mathbf{k}-%
\mathbf{k^{\prime }}$ is the momentum change due to  spin flip
scattering by fluctuations in SO field. Since the system is
macroscopically  isotropic in the $xy$ plane, the 
coordinate-independent  function $S_{\mathbf{k}}$ depends
only on $k$, that yields $S_{k}=S_{k^{\prime }}$,  and thus we
obtain
\begin{equation}
\label{11}
\mathrm{St}\,\rho _{\overline{\mathbf{k}}}=
-\frac{S_{k}\sigma_{z}}{\tau_{k}^{s}},
\end{equation}
where $\tau_{k}^{s}$ is the spin relaxation time,
\begin{equation}
\label{12}
\frac{1}{\tau_{k}^{s}}=\frac{m}{4\pi }
\int_{0}^{2k}C_{ \lambda  \lambda }(\mathbf{q})
\left( 4k^{2}-q^{2}\right)^{1/2}dq .
\end{equation}
We employ the following form of
$C_{\lambda\lambda}(\mathbf{q})$:
\begin{equation}
\label{13}
C_{ \lambda  \lambda }(\mathbf{q})=
2\pi \left\langle \lambda^{2}\right\rangle R^{2}\, e^{-qR},
\end{equation}
where $R$ is the length scale of variations
in $\lambda$. This form of correlator is realized when the Rashba
SO coupling is formed by the $z$-component of electric field of
random donors \cite{Ando} symmetrically distributed on
both sides of the QW at the distance $L=R/2$ from the
QW symmetry plane.

With Eqs.~(\ref{12}) and (\ref{13}) we obtain
\begin{eqnarray}
\label{15}
\frac{1}{\tau_{k}^{s}}
=\frac{1}{2\tau_{s0}}
\int_{0}^{2kR}e^{-x}\left( 4k^{2}R^{2}-x^{2}\right)^{1/2}dx
\nonumber  \\
=\frac{\pi Rk}{2\tau_{s0}}
\left[I_{1}(2kR)-L_{1}(2kR)\right],
\end{eqnarray}
where $I_{1}(x)$ and $L_{1}(x)$ are the Bessel and Struve functions,
respectively, and  $1/\tau_{s0}\equiv m\left< \lambda^2\right>$.
As a result, we obtain
\begin{eqnarray}
\label{16}
\frac{1}{\tau_{k}^{s}}=\frac{1}{\tau_{s0}}
\times \left\{
\begin{array}{ll}
kR, & kR\gg 1, \\
\\
\pi (kR)^{2}/2, & kR\ll 1. 
\end{array}
\right.
\end{eqnarray}

Equation (\ref{16}) agrees with the results of
Ref.[\onlinecite{Glazov05}] for $ kR\gg 1$, and shows that 
 for given $\tau_{s0}$  the relaxation rate rapidly decreases at small $kR$. 
 Due to the anticommutator form of $H_{\rm so},$ at small $kR$ 
main contribution to the SO field comes from 
the derivatives of $\lambda({\bf r})$. 
The increase in the relaxation time in this regime 
can be understood as a decrease in the disorder effect due to the
averaging of $H_{\rm so}$ over the area of $1/k^2$. The
dependence of the spin relaxation time on the correlation radius
$R$ is presented in Fig.~1.

{\it Combined resonance.} Now we consider response of the system
in a static magnetic field ${\mathbf B}$ to an external periodic field
$\mathbf{A}(t)=\mathbf{A}_{0}e^{-i\omega t}$. The corresponding
interaction $V_{\mathrm{ext}}=-(e/c)\hat{\bf v}{\bf A}$, where
${\bf v}=i[H_{0}+H_{\mathrm{so}},{\bf r}]$ is the velocity
operator, induces combined resonance causing transfers between
states with different spins and momenta. To study the spin
dynamics, we retain only the spin-related part of the
Hamiltonian and present it as:
\begin{eqnarray}
\label{17}
&&\hspace{-0.9cm}H=H_{0}+H_{\mathrm{so}}+H_{B}+V_{\mathrm{ext}},  \label{periodic} \\
&&\hspace{-0.9cm}H_{B}=\frac{\Delta}{2}({{\bm \sigma}}\cdot {\mathbf n}),\quad
V_{\mathrm{ext}}=-\lambda(\mathbf{r})\frac{e}{c}
\left( \sigma_{x}A_{y}-\sigma _{y}A_{x}\right),
\end{eqnarray}
where $\Delta=g\mu_{B}B$, $g$ is the electron Lande factor, and ${\mathbf n}$ is the direction of
${\mathbf B}$.  We include the magnetic field via the
Zeeman term, while neglect its orbital effects. The electron
energy spectrum is then spin-split,
$\varepsilon_{\mathbf{k}\uparrow ,\downarrow }
= \varepsilon _{\mathbf{k}}\pm \Delta /2$, where arrows correspond
to the direction parallel and opposite to the magnetic field. 
At realistic conditions, the splitting is much smaller than the
chemical potential $\mu$ of the degenerate electron gas.  Here the
periodic field leads to a disorder in $V_{\rm ext}$ due to the
factor $\lambda({\bf r})$. As a result, $V_{\rm ext}$ causes transitions
with the change in the electron momentum and spin in a single process, as shown in Fig.2. 
This is in contrast to conductivity, where the coupling 
of the external field to the disorder appears only through the
disorder effect on the electron states, and the transitions are
momentum-conserving. We calculate below the corresponding  spin pumping rate.

\begin{figure}[tbp]
\vspace*{-0.2cm} \epsfxsize=4.0cm \epsfbox{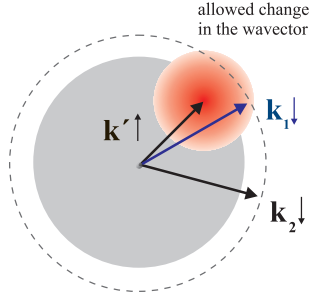}
\vspace*{-0.2cm} \caption{(Color online) Possible spin-flip
transitions. The initial state is
$\overline{\mathbf{k}}\mathbf{^{\prime }}\uparrow$, the final
states are $\overline{\mathbf{k}}_{1,2}\downarrow$. Dashed circle
corresponds to the transitions allowed by the energy conservation.
The region in the momentum space with the size of the order of
inverse correlation length $1/R$, where the transitions can occur
effectively, is also marked in the figure. If the final state is
inside this area, the transition has a relatively high
probability.}
\end{figure}

For given geometry of the external fields, denoted here as $[g]$
the time evolution of the spin projected electron density
$n_{\sigma}^{[g]}$ is due to the spin-gain 
$I_{\sigma ^{\prime }\rightarrow \sigma}^{[g]}(\omega,\Delta)$ and spin-loss 
$I_{\sigma\rightarrow\sigma ^{\prime }}^{[g]}(\omega ,\Delta )$ processes:
\begin{equation}
\label{21}
\frac{dn_{\sigma }^{[g]}}{dt}=I_{\sigma ^{\prime }\rightarrow \sigma }^{[g]}(\omega
,\Delta )-I_{\sigma \rightarrow \sigma ^{\prime }}^{[g]}(\omega ,\Delta ).
\end{equation}
The concentration gain
\begin{eqnarray}
\label{22}
&&I_{\sigma ^{\prime }\rightarrow \sigma }^{[g]}(\omega ,\Delta )
={2\pi}\sum_{\overline{\mathbf{k}}\overline{\mathbf{k}}\mathbf{^{\prime }}}
\left|
\left<\sigma\right|
W_{\overline{\mathbf{k}}\overline{\mathbf{k}}\mathbf{^{\prime}}}^{[g]}
\left|\sigma ^{\prime }\right>
\right| ^{2}
\nonumber \\
&&\times \left[ f(\varepsilon _{\overline{\mathbf{k}}\mathbf{^{\prime }}%
})-f(\varepsilon _{\overline{\mathbf{k}}})\right]
\delta (\varepsilon _{\overline{\mathbf{k}}}
-\varepsilon _{\overline{\mathbf{k}}\mathbf{^{\prime }}}-\omega),
\end{eqnarray}
is due to all possible transitions from occupied $%
\sigma ^{\prime }$ to unoccupied $\sigma $ states;
a similar expression holds for the loss $I_{\sigma \rightarrow \sigma ^{\prime }}(\omega ,\Delta )$.
The perturbation associated with the dipole moment acquired by electron spin in the
presence of SO coupling \cite{Rashba62,Rashba03} has the
form
\begin{equation}
W_{\overline{\mathbf{k}}\overline{\mathbf{k}}\mathbf{^{\prime }}}^{[g]}=
\frac{e}{c} \lambda_{\mathbf{k-k}^{\prime }}
\left(\sigma_{y}A_{0x}-\sigma_{x}A_{0y}\right).
\label{23}
\end{equation}
Due to charge
conservation, $d(n_{\sigma}^{[g]}+n_{\sigma ^{\prime }}^{[g]})/dt=0$.  Thus,
pumping rate  for the spin density component along the 
magnetic field $\left<s_{\bf B}^{[g]}\right>$, is 
\begin{equation}
\label{24}
\frac{d\left<s_{\bf B}^{[g]}\right>}{dt}=
\frac{d}{dt}
\frac{n_{\uparrow}^{[g]}-n_{\downarrow}^{[g]}}{2}
=\frac{dn_{\uparrow}^{[g]}}{dt}.
\end{equation}

\begin{figure}[tbp]
\vspace*{-0.2cm} \epsfxsize=4.2cm \epsfbox{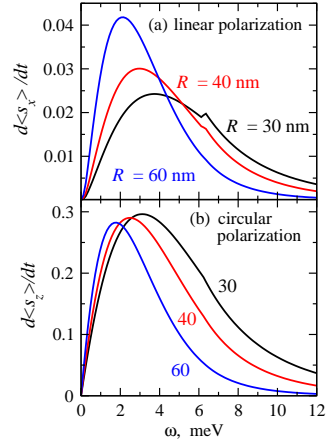}
\caption{(Color online) Spin pumping as a
function of $\omega $ for different values of correlation radius
$R$ marked near the lines. (a) In-plane field and 
linearly polarized radiation. Weak peaks seen at small $R$ at $\omega=\mu$ 
are manifestations of the spin-split density of states.  We use units
$(d\left<s\right>/dt)_0={1}/{\pi}\times({e^{2}}/{c^{2}})A_{0x}^{2}/{\tau_{s0}}$.
(b) $z$-axis field with $B\rightarrow0$ and circularly polarized radiation. We use units
$(d\left<s\right>/dt)_0={2}/{\pi}\times({e^{2}}/{c^{2}})A_{0}^{2}/{\tau_{s0}}$.
}
\end{figure}

Substituting (\ref{23}) into (\ref{22}) and averaging over spin disorder, we
obtain the component of the generation rate
\begin{eqnarray}
\label{25}
&&I_{\sigma ^{\prime}\rightarrow \sigma }(\omega,\Delta) =
2\pi \frac{e^{2}}{c^{2}}
K_{\sigma ^{\prime}\rightarrow \sigma }^{[g]}
\sum_{\mathbf{kk^{\prime}}}C_{ \lambda  \lambda }(\mathbf{q})
\nonumber  \\
&&\times\left[f(\varepsilon_{\overline{\mathbf{k}}^{\prime}})
-f(\varepsilon_{\overline{\mathbf{k}}^{\prime }}
+\omega )\right]\delta(\varepsilon_{\overline{\mathbf{k}}}
-\varepsilon_{\overline{\mathbf{k}}^{\prime}}-\omega ),
\end{eqnarray}
where the coefficient $K_{\sigma ^{\prime}\rightarrow \sigma }^{[g]}$ is determined by the 
field configuration. Here we consider two geometries:
(i) In-plane magnetic field ${\bf B}=(B,0,0)$, $\left<s_{\bf B}^{[g]}\right>=\left<s_x\right>$,  
linearly polarized radiation ${\bf A}_0=(A_{0x},A_{0y})$
and (ii) $z$-axis magnetic field ${\bf B}=(0,0,B)$,  $\left<s_{\bf B}^{[g]}\right>=\left<s_z\right>$, 
circularly polarized 
radiation ${\bf A}_0=A_{0}({\bf x},i {\bf y})$. In the case (i)
$K_{\sigma ^{\prime}\rightarrow \sigma }^{[g]}=A_{0x}^{2}$.
The transition rates satisfy symmetry relation $I_{\sigma ^{\prime}\rightarrow \sigma }(\omega ,\Delta )
=I_{\sigma \rightarrow \sigma ^{\prime}}(\omega ,-\Delta )$, and 
${dn_{\sigma }^{[g]}}/{dt}$ in Eq.(\ref{21}) vanishes if $B=0$.
In the case (ii) $K_{\sigma ^{\prime}\rightarrow \sigma }^{[g]}=0$
for the transitions up-down and $K_{\sigma ^{\prime}\rightarrow \sigma }^{[g]}=2A_{0}^{2}$
for the opposite process. Therefore, for the circularly polarized radiation 
the spin pumping occurs even at $B\rightarrow0$. 

The exact formula for the
pumping rate, valid in the general case $\mu
>\Delta/2$, can be obtained after integrating Eq.(\ref{25}) over the directions of
$\mathbf{q}=\mathbf{k}-\mathbf{k^{\prime }}$. As a result one
obtains (with $y=kR$)
\begin{eqnarray}
\label{27}
I_{\sigma ^{\prime}\rightarrow \sigma }(\omega,\Delta) &=&
\frac{2}{\pi}\frac{e^{2}}{c^{2}}
K_{\sigma ^{\prime}\rightarrow \sigma }^{[g]}
\frac{1}{\tau_{s0}}
\int_{y_{{\rm min},\sigma^{\prime}}}^{y_{{\rm max},\sigma ^{\prime}}}y\,dy
\\
&&\hspace{-2cm}
\times
\int_{|y-y_{\sigma^{\prime}}|}^{y+y_{\sigma^{\prime}}}
\frac{e^{-x}x\,dx}{[4y^{2}x^{2}-(x^{2}+2m(\sigma^{\prime}\Delta -\omega)R^{2})^{2}]^{1/2}},\nonumber
\end{eqnarray}
where $y_{{\rm min},\sigma^{\prime}}=R\left[\max\{0,2m(\mu -\sigma^{\prime}\Delta /2-\omega )\}\right]^{1/2}$,
$y_{{\rm max},\sigma ^{\prime}}=R[2m(\mu -\sigma ^{\prime}\Delta /2)]^{1/2}$,
$y_{\sigma^{\prime}}=[y^2+2m(\omega -\sigma ^{\prime}\Delta )R^2]^{1/2}$,
and $1/\tau_{s0}$ is the prefactor in Eq.~(\ref{15}).

For numerical calculations we use the following parameters: electron
effective mass for a Si/Ge (001) QW $m=0.19\, m_0$ (where $m_0$
is free electron mass), electron concentration per valley
$n=5\times 10^{11}$~cm, Fermi momentum $k_F=1.8\times
10^6$~cm$^{-1}$, and  the Fermi energy $\mu=6.3$~meV. To calculate 
$d\left<s_x\right>/dt$ we take magnetic field $B=1$~T leading to the spin splitting
$\Delta=0.12$~meV for $g=2$. 
The injection rates $d\left<s_{\bf B}^{[g]}\right>/dt$ are
presented in Fig.~3. For both radiation polarizations, the 
peaks have the maximum position and the width on the order of $\omega \sim \mu /k_{F}R.$
With the increase in $R$ at given $k_F$, the available momentum and energy ranges decrease, 
the peaks sharpen and shift to lower frequencies. 
For linearly polarized radiation (Fig.~3(a)), the pumping rate is linear in $\Delta$.

{\it Conclusions} We have studied spin relaxation and infrared
radiation-induced spin transitions in a
2D electron gas with the Rashba field disorder. Quantum effects
related to noncommutativity of the momentum and random Rashba
potential lead to the decrease in the spin relaxation rate
when the spatial scale of the randomness is smaller than the
electron wavelength. In contrast to conductivity, external
periodic electromagnetic field generates a perturbation directly 
including the SO disorder, and, therefore, causing
spin-flip accompanied by a momentum change. 
As a result, electron spin density can be pumped by
coupling of spins to the external periodic field in the frequency range 
up to the Fermi energy. These effects extend the abilities 
of manipulating spins in semiconductor structures.

{\it Acknowledgements.}
This work is partly supported by the FCT Grant PTDC/FIS/70843/2006 in
Portugal and by Polish Ministry of Science and Higher Education as a
research project in years 2007 -- 2010. E. Sherman acknowledges support
of the University of Basque Country UPV-EHU grant
GIU07/40 and valuable discussion with M.M. Glazov.

\end{document}